%% file: main.tex
\documentclass{llncs}

\usepackage[utf8]{inputenc} 
\usepackage[T1]{fontenc}   
\usepackage{hyperref}      
\usepackage{url}           
\usepackage{booktabs}      
\usepackage{amsfonts}       
\usepackage{nicefrac}    
\usepackage{microtype}     
\usepackage{graphicx}
\usepackage[numbers]{natbib} 

\usepackage{graphicx}
\title{A General Framework for Revealing Human Mind with auto-encoding GANs}
\author{Pan Wang\inst{1}\and
Rui Zhou\inst{1}\and
Shuo Wang\inst{1}\and Ling Li\inst{2}\and Wenjia Bai\inst{1}\and Jialu Fan\inst{1}\and Chunlin Li\inst{3}\and Peter Childs\inst{1}\and Yike Guo\inst{1,4}}
\authorrunning{F. Author et al.}
\institute{Imperial College London, United Kingdom \and
University of Kent, United Kingdom\and Beijing Institute of Technology, China\and
Hong Kong Baptist University, Hong Kong\\}
\begin{document}
\maketitle
\begin{abstract}
  ‘How to visualise human mind?’ is a key question in brain decoding. Addressing this question could help us to find regions that are associated with observed cognition and responsible for expressing the elusive mental image, leading to a better understanding of cognitive function. The traditional approach treats brain decoding as a classification problem, reading the mind through statistical analysis of brain activity. However, human thought is rich and varied, that it is often influenced by more of a combination of object features than a specific type of category. For this reason, we propose an end-to-end brain decoding framework which translates brain activity into an image by latent space alignment. To find the correspondence from brain signal features to image features, we embedded them into two latent spaces with modality-specific encoders and then aligned the two spaces by minimising the distance between paired latent representations. The proposed framework was trained by simultaneous electroencephalogram (EEG) and functional MRI (fMRI) data, which were recorded when the subjects were viewing or imagining a set of image stimuli. In this paper, we focused on implementing the fMRI experiment. Our experimental results demonstrated the feasibility of translating brain activity to an image. The reconstructed image matches image stimuli approximate in both shape and colour. Our framework provides a promising direction for building a direct visualisation to reveal human mind.
\end{abstract}
\section{Introduction}

‘Mind reading’ technology belongs to the traditional neuroscience field of ‘brain decoding’\cite{naselaris2011encoding}\cite{kamitani2005decoding}\cite{haynes2006neuroimaging}\cite{tong2012decoding}, which attempts to reconstruct sensory stimuli from information already encoded in the brain using measured neuronal activity. The use of classification through statistical analysis of brain activity and its corresponding stimuli is well-established in the field of brain decoding; many studies have demonstrated that visual features\cite{kamitani2005decoding}, such as object categories\cite{cox2003functional}\cite{haxby2001distributed}, shape dimensions and orientations\cite{georgieva2009processing}\cite{kourtzi2003integration}, can be decoded from brain activity. Recently, brain decoding based on deep learning methods have been proposed by researchers such as Horikawa et al.\cite{horikawa2017generic}, who demonstrated a strong correlation between visual cortical activity and CNN visual features by decoding brain activity into hierarchical visual features of a pre-trained deep neural network from the same image. 
However, these research are mainly focused on solving a classification problem, which is the concept level decoding limited by classes in the training. Also, the sensitiveness to the training algorithm resulting in poor precision so far. Latest research have shown the potential breaking these limitations from brain decoding with deep learning method. Based on the work done by Horikawa et al., Shen et al.\cite{shen2019deep}used an iterative method to optimise the pixel values of DNN features of the current image to reconstruct the corresponding image. A recent research published in Nature demonstrated direct synthesis of brain signal into a sentence, translating human brain activity into language that we can understand\cite{anumanchipalli2019speech}. Human vision is composed of multiple visual features. Due to the diversity and complexity of the human visual world, our mind may compress a combination of features as a new image which does not belong to any existing categories. For example, in bionic design, a designer needs to extract features from both biological domain and product domain before combining them in mind to create assembled features and products. Therefore, a simple classification question may be inadequate, being able to decode human thought through  features(brain signal and image) mapping will give more meaning for reconstructing human thought.
\begin{figure}
    \centering
    \includegraphics[width=3.5in]{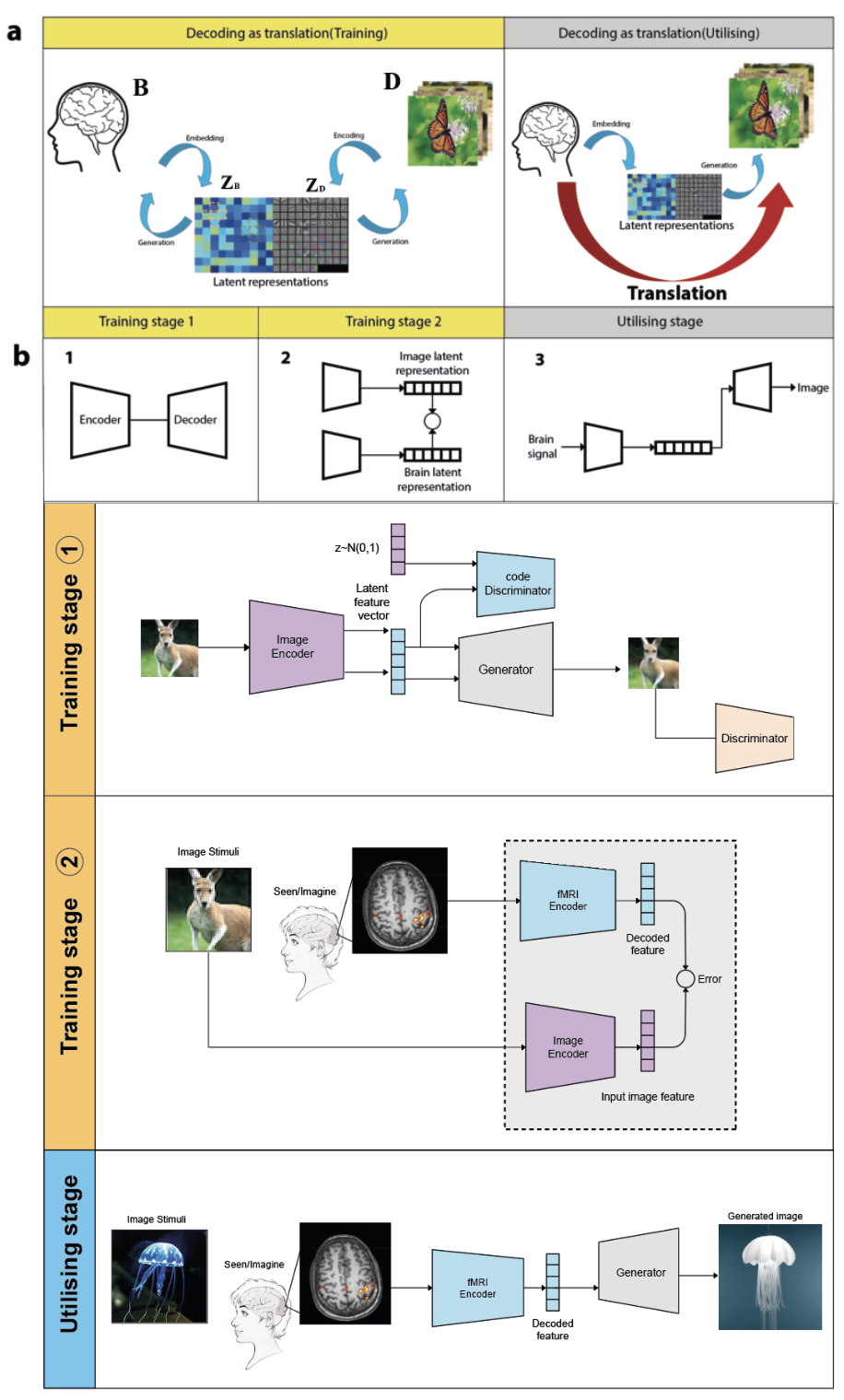}
    \caption{Decoding as translation. (a)Brain-to-image representational paradigm. Translating brain activity into image through embedding both image and brain activity into latent representations, then minimising the distance between the two latent representations for image reconstruction. (b)Training process and main model architecture. This framework contains three stages: two training stages,1)pre-training a variational auto-encoder for image feature learning and an image generator;2) training brain signal feature encoder which map brain signal feature to image feature. One utilising stage, inputting brain signal to the model then output the corresponding image.}
    \label{fig:my_label}
\end{figure}

Decoding as translation. Our goal was to demonstrate the feasibility of decoding human thought by translating brain activity into an image. We proposed to establish an end-to-end mental image generation framework, in which the mental image generation can be done through mapping brain signal feature to image feature via latent representations, shown in figure 1. We embedded both image and brain activity into two latent representations, then we mapped the two latent representations by minimising the distance in between. In the end, the mental image was generated from the mapped latent representations. 

To establish such a framework, an EEG and fMRI simultaneous recording experiment have been designed for brain data collection. The whole experiment contained two sessions, an image presentation session and a mental imagery session. The image presentation session was used for model building and training, which established the correlations between the seen image and brain activity measured through EEG and fMRI. The mental imagery session was used for reconstructing the mental image, which required the subject to do the imagery task under instructions. As shown in figure 2, we build an end-to-end model that could translate brain activity A to a mental image B. During the training, 1) we train a variational auto-encoder which extracts the image feature vector to a latent representation and an image generator. 2) A brain signal encoder was trained to map brain activity to its latent representation. 3) Then we generate the mental image from both image presentation experiment and imagery experiment. In addition, compared to previous research which used different models to implement different brain signals, we built a common framework for revealing the human mind from both EEG and fMRI signals. 

The main contributions of this work are listed as follows.
We proposed an end-to-end mental image reconstruction framework which reveals the human mind by a direct translation from brain activity into images through both EEG and fMRI signals. The proposed generative brain decoding approach generates the mental image by mapping features from brain activity to image via a latent representation.
We contributed an EEG and fMRI simultaneous recording dataset for revealing the human mind, which also supports that our framework is suitable for input from different brain signals. We plan to publish this dataset upon the acceptance of the paper.

\input{problem/problem_formulation.tex}

\input{Method/method.tex}

\input{experiment/experiment.tex}

\input{result/result.tex}

\bibliographystyle{plain}
\bibliography{ref}

\end{document}

%% file: problem/problem_formulation.tex
\label{gen_inst}

%% file: Method/method.tex
\section{Method}
\label{headings}
\subsection{Overview and Notations}
This framework takes either EEG or fMRI signal as input and outputs the reconstructed corresponding seen images. This framework includes an EEG signal encoder $E_{e}(.)$, an fMRI encoder $E_{f}(.)$ and an image re-constructor $R_{i}(.)$. 
Our training process include three phases. 
Firstly, we train a variational autoencoder, which contain an image encoder $E_{i}(.)$ and an image re-constructor $R_{i}(.)$. The image encoder encodes an image $i$ to a feature representation vector $z_i = E_{i}(i)$. The image re-constructor reconstructs the image $i'$ based on $z_i$. Secondly, we train an EEG signal encoder $E_{e}(.)$ and an fMRI signal encoder $E_{f}(.)$. The EEG signal encoder maps the EEG signal $e$ to the image feature representation vector $z_{ei}$. The fMRI signal encoder maps the fMRI signal $f$ to the image feature representation vector $z_{fi}$. The image feature representation vectors are learned by the image encoder pre-trained in the first phase.
Thirdly, we combine the trained EEG signal encoder, the trained fMRI signal encoder and the trained image re-constructor to construct the final image reconstruction framework. When the framework receives an EEG signal or a fMRI signal, the signals will be encoded by the EEG signal encoder and fMRI signal encoder into the image feature representation vectors $z_{ie}$ and $z_{if}$. Then the latent representation $z_{ie}$ or $z_{if}$ will be fed into the image re-constructor to reconstruct the seen image $i'$. In general, this framework has two main tasks including learning the feature of brain signal; reconstructing the image based on the learned feature. In this section, we present the models designed to solve those two tasks.
\begin{figure}
\centering
\includegraphics[width=0.8\textwidth]{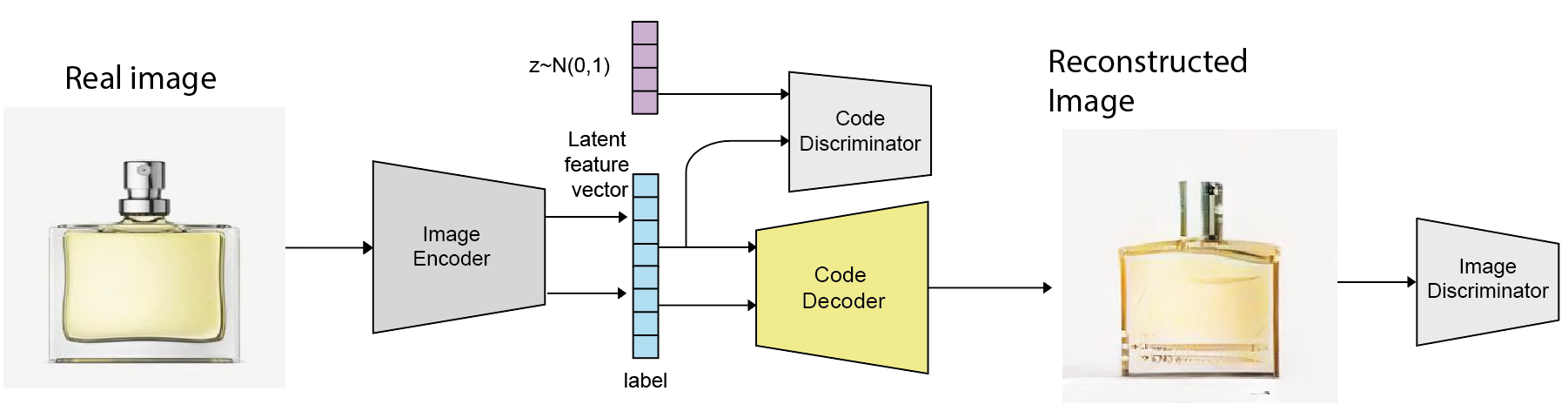}
\caption{Diagram of the proposed $\alpha$-GAN model.} \label{fig1}
\end{figure}
\subsection{Image feature learning \& image reconstruction}
To solve the tasks of reconstructing the image based on the learned feature, we firstly train a variational auto-encoder. Specifically, we employ the $\alpha$-GAN structure \cite{rosca2017variational} to construct this variational auto-encoder. The main feature of $\alpha$-GAN is that it combines standard GAN structure with variational auto-encoder.  The advantage of this structure is that it resolves the blurriness problem of traditional variaional auto encoders by employing adversarial loss in GAN part, and also fixes GAN's model collapse problem via reconstruction loss from the variational auto-encoder part.  The $\alpha$-GAN consists of four networks: an image encoder $E_{i}(.)$, an image re-constructor $R_{i}(.)$, a code discriminator $C_w(.)$ and an image discriminator $D_{\phi}(.)$. As all the variational auto-encoders, the $\alpha$-GAN optimises its weight trough maximising the evident lower bound (ELBO). The first part of the ELBO - the likelihood is decomposed as the combination of a reconstruction loss and an adversarial loss introduced by the image discriminator. The second part of the ELBO - the Kullback–Leibler divergence can be represented as the adversarial loss introduced by the code discriminator. Thus, the loss function of $\alpha$-GAN can be expressed as:
\begin{equation}
   L(\theta, \eta) = \mathbb{E}_{q_{\eta}(z|x)}[
   - \lambda ||x - R_{i}(z)||_1 + \log \frac{D_{\phi}(R_{i}(z))}{1 - D_{\phi}(R_{i}(z))} + \log \frac{C_w(z)}{1 - C_w(z)}].
\end{equation}
where $C_w(.)$ discriminates whether latent variable $z$ is produced by the encoder or the standard Gaussian distribution; $q_{\eta}(z|x)$ is the recognition distribution used to approximate the true posterior.

\subsection{Brain signal feature learning}
As our framework is designed to receive both EEG signal and fMRI signal, brain signal feature learning consists of two parts: an EEG signal encoder and an fMRI signal encoder.

\subsubsection{EEG signal encoder \& fMRI signal encoder}
There are a lot of studies contributing to the EEG signal feature extraction, e.g.,  \cite{palazzo2017generative} propose using recurrent neural networks to encode EEG signal; \cite{schirrmeister2017deep}, treat EEG signals as 2D signals like the grey image and using 2d convolutional neural networks 
to learn the feature of the EEG signal. Unlike the EEG signal which can be directly used to train the learning model, in the studies relevant to the fMRI signal, e.g., fMRI signal classification and image reconstruction based on fMRI signal, prepossessing needs to be done based on the fMRI signal before the signal can be used in model training.  

To construct our EEG signal encoder and fMRI signal encoder, we employ the same structure. We firstly employ Pearson Correlation Coefficient to select $k$ most related features of the EEG signal or the fMRI signal according to each dimension of the $m$ dimension image feature representation vector. Then, we construct $m$ parallel Bayesian regression sub-models. Each Bayesian regression predicts the value of one corresponding dimension of the image feature representation vector based on $k$ most related features of the EEG signal or the fMRI signal. Thus the equation of this encoder can be presented as: 
\begin{equation}
    z_{i}^j = E_{f}^{j}(f^{j})
\end{equation}
or 
\begin{equation}
    z_{i}^j = E_{e}^{j}(e^{j}),
\end{equation}
where $(j = 1,... ,m)$ represent the $j^{th}$ dimension of the image feature representation vector. 

\begin{figure}
\centering
\includegraphics[width=1.0\textwidth]{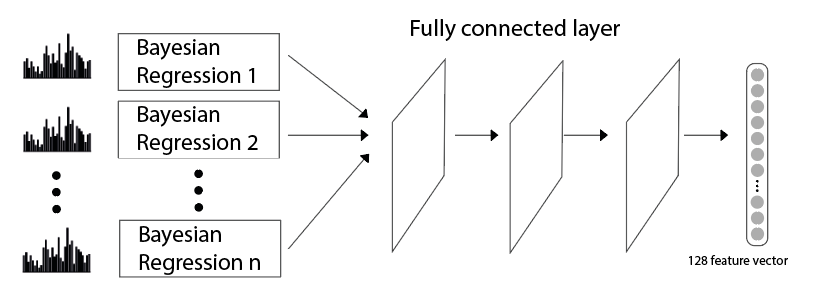}
\caption{Architecture of the EEG and fMRI feature encoder.} \label{fig1}
\end{figure}

%% file: experiment/experiment.tex
\section{Experiments}
\label{others}

Our experiment contains two parts, a simultaneous EEG and fMRI recording experiment for collecting brain activity, and an end-to-end model training experiment for optimising the brain translation framework, which translates the recorded brain activities into images.

\subsection{Dataset acquisition}
\subsubsection{Subjects and equipments}
The data sets were collected from five healthy right handed subjects, aged 25-27 years, with normal or corrected-to-normal vision. All subjects were highly trained for participating in EEG and fMRI experiments. Consent forms had been signed by all participants in the experiment and the study protocol was approved by the Research Ethics Committee. 

The fMRI data were collected using a 3.0-Tesla Siemens MAGNETOM Prism scanner. To acquire full-brain functional images, an interleaved T2-weighted gradient-echo echo planar imaging (EPI) scan was used (TR, 2000 ms; TE, 30ms; flip angle, 90 deg; FOV, 64 $\times$ 100 mm; voxel size, 3.5 $\times$ 3.5 $\times$ 3.5 mm; slice gap, 3.5mm; number of slices, 33; multiband factor,0). Anatomical images in high resolution, typically for full-brain T1-weighted magnetization-prepared rapid acquisition gradient-echo (MP-RAGE) fine-structural images, were also acquired (TR, 2300 ms; TE, 2.36 ms; TI, 900 ms; flip angle, 8 deg, FOV, 256 $\times$ 256 mm; voxel size, 0.84 $\times$ 0.84 $\times$ 1.0 mm).

The electroencephalogram (EEG) was recorded (band-pass 0.05-100 Hz, sampling rate 1000 Hz) from a set of 64 Ag/AgCl electrodes according to the 10–20 system with the Neuroscan Synamp2 Amplifier. 

The EEG electrodes were on-line referenced to the average of the left mastoid and then off-line referenced to the average of two mastoids. An electrode was applied to the cephalic location as the ground. Vertical electro-oculograms (EOG) were recorded with electrodes placed on the supra-orbital and infra-orbital locations of the left eye, whereas the horizontal electro-oculogram was recorded from electrodes on the outer canthi of both eyes. Furthermore, electrode impedance were set to stay below 5 k$\Omega$ throughout the whole experiment.

\subsubsection{Visual stimuli}

The visual stimuli used in the experiment design consisted of 20 category (eg.backpack, bulb, jelly fish, leaf et. al), each category contained 50 images. All images were collected online and were resized to 500x500 pixels which ensured the main contents of the image were at the centre with similar sizes. 

\subsubsection{Experiment sessions}
Two  types  of  experiments  were  carried  out  in  the  brain  data  collection:  an  image
presentation experiment and an mental imagery experiment. The image presentation experiment consisted of five sub-sessions, in which images from 20 different categories were presented. The data were collected altogether in order to have a flexible split into training and testing sets during the model training. Subjects were asked to fixate on a white cross on the screen, which they viewed through a mirror attached to the head coil. All the visual stimuli were rear-projected onto a translucent screen in the fMRI scanner via a liquid crystal display projector. Data from each subject were collected through several scanning sessions spanning approximately 2 months. Only one of the two experiments was conducted on one subject on each experimental day.

The image presentation experiment consisted of five sub-sessions, the images were randomly selected from the total set of 1,000 images (50 images each for 20 categories). Each sub-session consisted of 200 unique images, and another 20 extra image trials from the same sub-session for repetition. Furthermore, during the experiment, the presentation order of the images was randomized across sub-sessions in order to minimise the confounding factors. Within each sub-session, images were presented at the centre of the display for 7.5 s. Between two consecutive images, a black screen was presented for 0.5 s in order to indicate the onset of each trial. Meanwhile, a trigger was marked onto the EEG recording for a short period of 0.01 s to separate out epochs. A complete block contained 10 images and 1 randomly selected image from the current run to repeat. Therefore, each sub-session had a total of 20 blocks (1 min 28 s, for each block). During the experiment, subjects maintained steady fixation throughout each sub-session and performed feedback responded with a button press for each repeated image in the block. This repetition detection task was designed to ensure their attention was retained on the presented images. The whole sub-session was set to be 1,760 s (29 min 20s), with a 5-min rest period between each sub-session. The total time for the experiment lasted for 10,000 s (2 hour 46 min 40 s).

\subsubsection{Data pre-processing (EEG $\&$ fMRI)}
All fMRI scans were pre-processed using SPM12 toolbox (https://www.fil.ion.ucl.ac.uk/spm/). Firstly, to correct for timing difference, slice timing was performed for the interleaved acquiring order in EPI scans. fMRI scans with head motion exceeding 3 mm were discarded to avoid large deviations in the data, before the three-dimensional motion correction was applied to the scans. Co-registration was then used to realign the fMRI scans to the high-resolution T1-weighted structural MRI images as the reference. After that, to minimise cross-body difference, the co-registered data was normalized following the registration between T1-weighted sMRI images and the MNI template. The normalization procedure in SPM12 by default included segmentation, bias correction, and spatial normalization. Finally, the normalized data were re-interpolated to a voxel size of 2 $\times$ 2 $\times$ 2 mm. 

EEG data was pre-processed using the EEGLAB toolbox
(https://sccn.ucsd.edu/eeglab/index.php). Viewing all channels with a low gain helped discard any bad channels. Given the triggers were marked simultaneously with the presentation of the images, i.e. the start of each trial, epoch extraction was applied to cut out the first 8 s of data after all triggers, with baseline correction to remove the effect of pre-stimulus data. In order to match the 4 segments of fMRI data in each trial, the EEG data was also cut into 4 segments of 2s. The data was also filtered using a basic FIR filter with a pass band from 0.05 to 100 Hz.

\subsection{Implementation details}
\subsubsection{Decoder training}
During the first phase, we train the decoder with $\alpha$-GAN structure which includes four networks. The image data set used to train this decoder includes 18 categories and each category has 1000 images. The training parameters of the decoder is as follows: the batch size is 18, the Adam optimisation algorithm is used to optimise the weights, the learning rate is $8 * 10^{-4}$ and the epsilon value is $1 * 10^{-8}$. For every row, we update the weights of the image encoder and the generator twice and then update the weights of the code discriminator and the image discriminator once. In order to let $\alpha$-GAN have good performance, we employ Residual block to construct our $\alpha$-GAN. Each Residual block has two 2D convolutional layers ($64$ channels, kernel size $3\times3$, stride size $1$ and padding size $1$), and each 2D convolutional layer is followed by a Relu activation layer and a batch normalisation layer. The image encoder contains one 2D convolutional layer ($64$ channels,  kernel size $4\times4$ and stride size $1$), three residual blocks and a fully connected layer ($65536\times128$). The 2D convolutional layer is followed by a Relu activation layer and an averaging pooling layer; each the first two residual blocks are follow by an averaging pooling layer. The image re-constructor has one fully connected layer ($65536\times128$), four residual blocks and one 2D convolutional layer ($3$ channels,  kernel size $3\times3$ and stride size $1$). The fully connected layer is followed by a Relu activation layer; Each residual block is followed by a bi-linear two-fold up-sampling layer and the 2D convolutional layer is followed by a Tanh activation layer. The code discriminator has three successive fully connected layers ($128\times700$, $700\times700$ and $700\times1$). The first two fully connected layers are followed by the leaky Rely activation layers; the last fully connected layer is followed by a Sigmoid activation layer. The image discriminator contains one 2D convolutional layer ($64$ channels,  kernel size $5\times5$, stride size $1$ and padding size $2$), three residual blocks and a fully connected layer ($65536\times1$). The 2D convolutional layer is followed by a leaky Relu activation layer and an averaging pooling layer; each the first two residual blocks are follow by an averaging pooling layer; the last fully connected layer is follow by a Sigmoid activation layer.

\subsubsection{fMRI and EEG encoder training}
During the second phase, we train the EEG signal encoder and the fMRI encoder. The EEG signals used to train the encoder come from 5 healthy volunteers. Each volunteer has watched 18 categories of images. Each EEG signal collected has 64 channels and lasts 500 milliseconds. Each fMRI signal is a 4500 dimensional vector which consists of all the voxels of the brain visual region. During the training, we select 500 the most related features of the EEG signal or the fMRI signal. As the image feature is a 128 dimensional vector, for either the EEG encoder or the fMRI encoder, we have 128 Bayesian regression models. 

\subsubsection{Evaluation metrics and baseline}
In order to quantitatively evaluate the quality of reconstructed images, we perform a pairwise structural similarity comparison experiment. In this experiment, the structural similarity index(SSIM)\cite{wang2004image} between the reconstructed image and the original image is compared with the SSIM between the reconstructed image and the images from other categories. The main reason that we adopt SSIM here is that it systematically compares the quality of the original image and reconstructed image by three comparisons including Luminance comparison, contrast comparison, and structure comparison. Specifically, in our case, after we reconstruct the images based on brain activity, we first calculate the SSIM $S_{Ori}$ between the reconstructed image and the original seen image. Then, as we have 18 categories of images, we randomly select one image from each of the rest of 17 categories of images and calculate the SSIM $S_{Non}$ between them and the reconstructed image. Finally, we compare the $S_{Ori}$ with the $S_{Non}$. If the $S_{Ori}$ is larger than the $S_{Non}$, we count this comparison as correct and add one to correct comparison number $n_{correct}$. For $n$ test reconstructed image, we do $17*n$ times comparison and the total correct rate is equal to $n_{correct}/(17*n)$. Besides the SSIM, we also calculate the Mean-Square-Error (MSE) and between two images. It is worth mentioning that, unlike the the SSIM comparison experiment which is correct when the $S_{Ori}$ is larger than the $S_{Non}$, the MSE comparison experiment is correct when $M_{Ori}$ is smaller than the $M_{Non}$.

From the \ref{fig:Bar}, we can learn that the correct rate of the reconstructed image for the $\alpha$-GAN is close to $100\%$. The correction rate of the reconstructed image for the EEG framework is very close to the correction rate of the reconstructed image for the fMRI framework, while the correction rate of the fMRI framework is higher than that of  the EEG framework. The main reason behind this result is cause by the different characteristics of their inputs. For the $\alpha$-GAN, its input is the real image which is clean date. For the fMRI framework, its input is the fMRI signal. Even the fMRI signal contains a lot of information, the size of voxel still limits the accuracy of reconstruction. For the EEG framework, its input is the EEG signal, which is noisier than the fMRI signal and contains less information.

%% file: result/result.tex
\section{Results and discussion}

\subsection{Seen image reconstruction}

\begin{figure}
    \centering
    \includegraphics[width=4.5in]{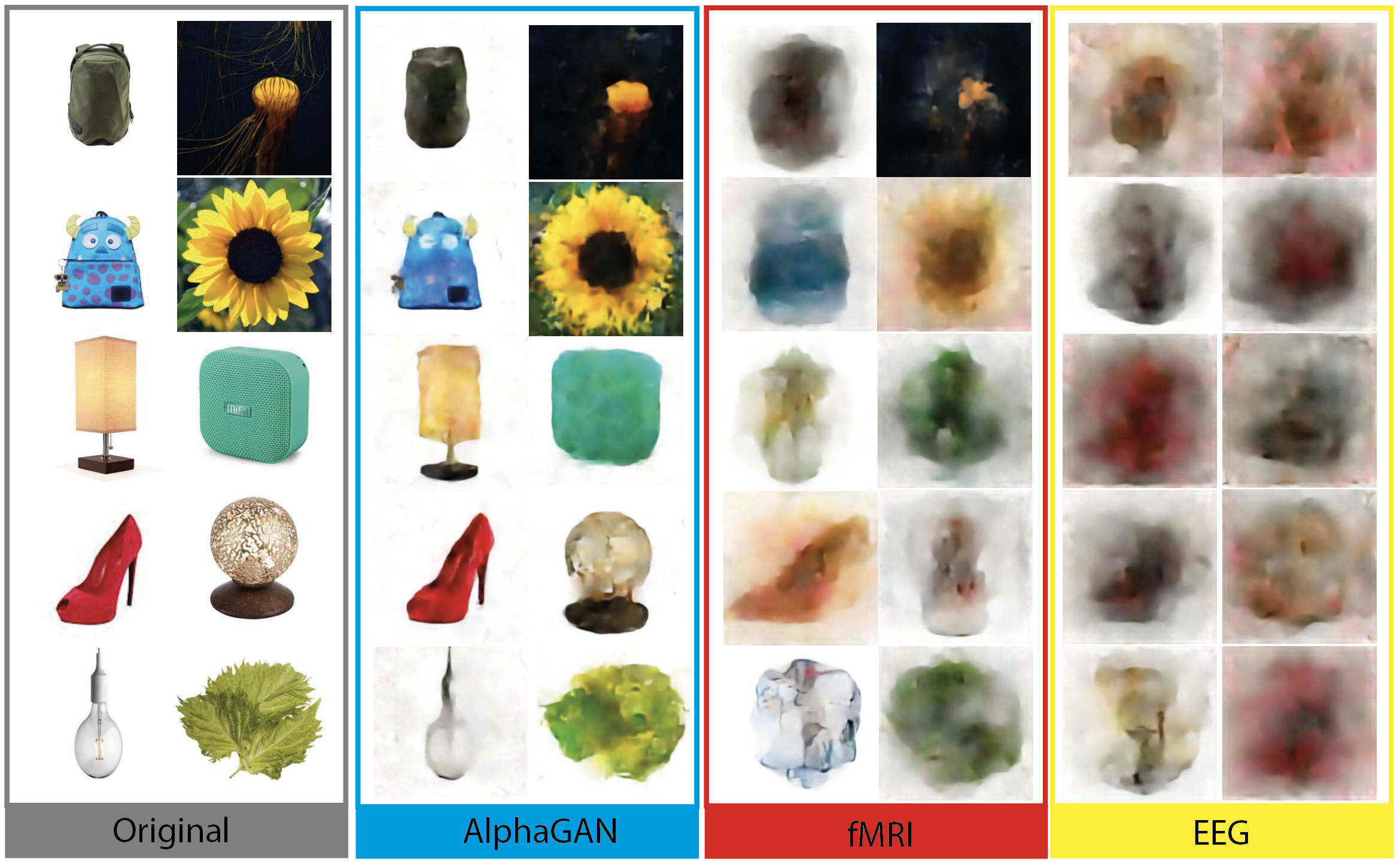}
    \caption{The reconstructed images based on the $\alpha$-GAN, the EEG framework and the fMRI framework.}
    \label{fig:IMAGE}
\end{figure}

\begin{figure}
    \centering
    \includegraphics[width=4.0in]{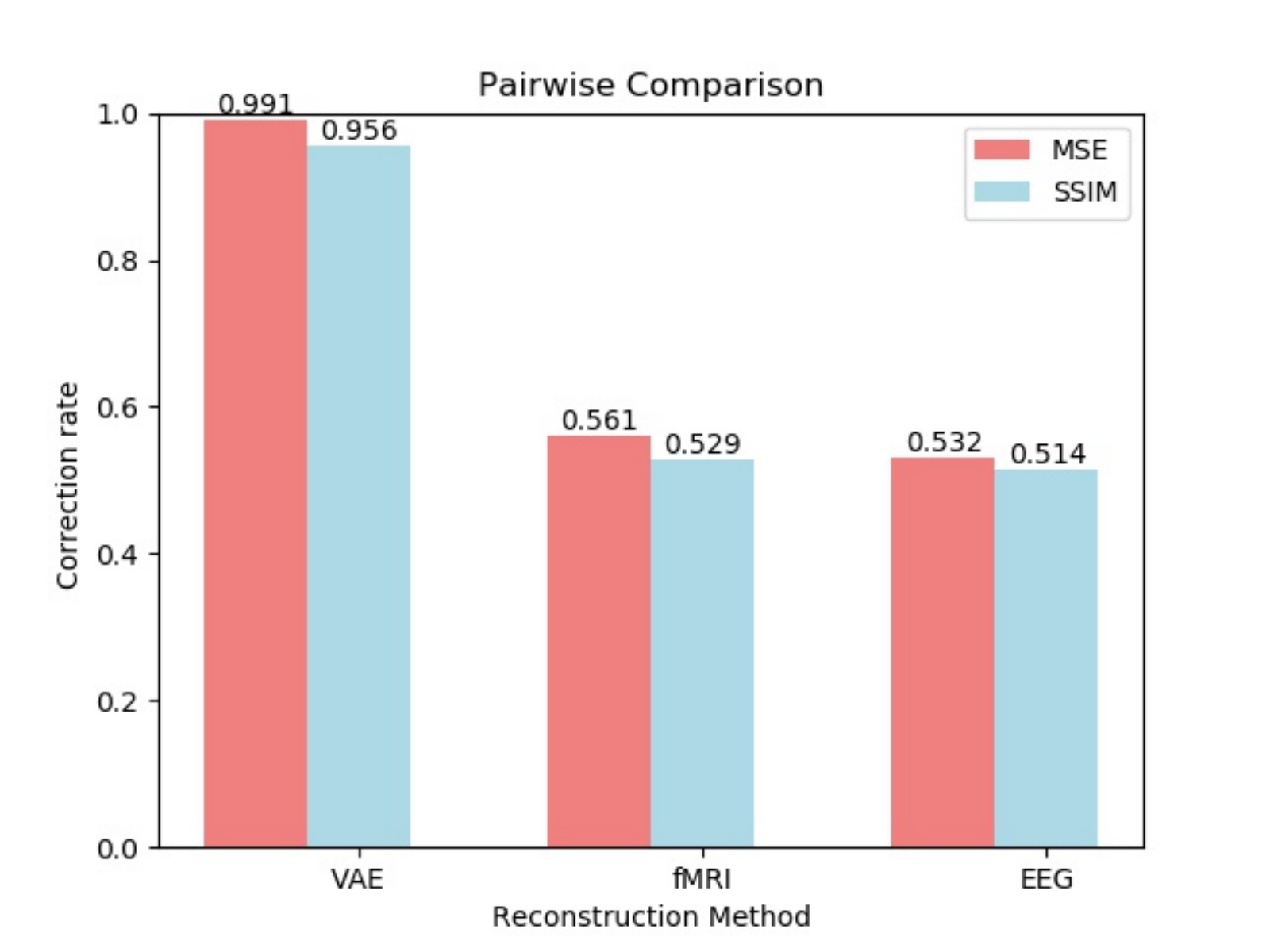}
    \caption{The pairwise comparison experiment for the reconstructed images based on the $\alpha$-GAN, the EEG framework and the fMRI framework.}
    \label{fig:Bar}
\end{figure}

Fig.\ref{fig:IMAGE} demonstrates the results of reconstructed images from image presentation session while subjects were required to view the presented images. Images shown in the gray, blue, red and yellow frame indicate the original image, generated images from AlphaGAN model, reconstructed fMRI mental images and reconstructed EEG mental images respectively. These results shown that our end-to-end mental image reconstruction framework could successfully reconstruct the seen image from the recorded brain signal(EEG \& fMRI). From the reconstructed images, we can learn that after 400 episodes, the $\alpha$-GAN model can reconstruct the clear images shown in blue frame. To evaluate the quality of reconstructions, both the SSIM comparison and the MSE comparison have been applied to the results shown in Fig.\ref{fig:Bar}. The correct rate can reach 100 percent, which means that the reconstructed image of $\alpha$-GAN model can keep most information from the original image. We could also see human mental image could be reconstructed successfully from the recorded fMRI signal, that we could distinguish both shape and colour of the mental content. However, mental image reconstruction from EEG signal is blurry, which is hard to distinguish with its shape. The SSIM comparison and MSE comparison have also been applied on fMRI and EEG rusults, the correct rate can still arrive around 60 percent, which is remarkable.

\subsubsection{EEG and fMRI comparison}
Comparing the reconstructed images based on the fMRI signal and the EEG signal, we can learn that the fMRI signal can reconstruct the image with higher quality which is within our expectation, as the fMRI signal contains less noise and has more information. The main limitation of the fMRI signal now is its voxel size which can be improved in the foreseeable future. Moreover, the SSIM comparison correct rate and the MSE comparison correct rate of fMRI framework can arrive around 80 percent. Considering the performance of EEG signal in our case, the EEG signal can hardly be applied to the seen image reconstruction task while its high correct rate means that the EEG signal maybe sufficient for the seen image classification task.

\section{Summary and Discussion}
In this paper, we have presented an end-to-end mental image reconstruction framework for translating human brain activity into an image. Our results show that mental content could be reconstructed form simultaneous recorded EEG and fMRI signal though the proposed framework. Differentiating with previous methods focus on classification, our method was certainly reconstruct mental image from feature level correspondence, which gives the possibility of visualising human thought without the limitation of classification. 
In addiction, we compared the mental image reconstruction ability from both fMRI and EEG signals, and found that fMRI signal contains more brain content information than EEG, even the classification rate of EEG signal is higher than fMRI.
\subsubsection*{Acknowledgments}

The authors would like to acknowledge Jaywing PLC who funded this research.